\documentclass[aps,prl,twocolumn,showpacs,preprintnumbers,amsmath,amssymb,floatfix]{revtex4}

\usepackage{graphicx}
\usepackage{dcolumn}
\usepackage{bm}
\newcommand{\cn}{\ensuremath{\mathcal{N}}}

\begin{document}


\title{High-flux beam source for cold, slow atoms or molecules}

\author{S.~E.~Maxwell}
\author{N.~Brahms}
\author{R.~deCarvalho}
\author{S.~V.~Nguyen}
\author{D.~Patterson}
\author{J.~M.~Doyle}
\affiliation{%
Department of Physics, Harvard University, Cambridge,
Massachusetts 02138, USA
}%

\author{J.~Helton}
\affiliation{%
Department of Physics, Massachusetts Institute of Technology, Cambridge,
Massachusetts 02139, USA
}%

\author{D.~R.~Glenn}
\author{J.~Petricka}
\author{D.~DeMille}
\affiliation{Department of Physics, Yale University, New Haven,
Connecticut 06520, USA
}%

\date{\today}

\begin{abstract}
We demonstrate and characterize a high-flux beam source for
cold, slow atoms or molecules. The desired species is vaporized
using laser ablation, then cooled by thermalization in a cryogenic
cell of buffer gas. The beam is formed by particles exiting a hole
in the buffer gas cell. We characterize the properties of the beam
(flux, forward velocity, temperature) for both an atom (Na) and a molecule (PbO) under
varying buffer gas density, and discuss conditions for optimizing
these beam parameters. Our source compares favorably to existing
techniques of beam formation, for a variety of applications.
\end{abstract}

\pacs{39.10.+j}
\maketitle

Cold, slow beams of atoms or molecules are of wide utility. A
common use of such beams is as a source for loading into traps,
where the particles can be further cooled and manipulated, e.g. to
create Bose condensates or Fermi degenerate gases. Because the
number of trapped particles is typically limited by the
characteristics of the initial beam (flux, forward velocity,
temperature, etc.), significant effort has been put into
developing improved atomic beam sources
\cite{lvis,haubeam,schoser:023410,dieckmann:3891,mm-mot,prentisssource,abrahamMagfilter}.
Work on developing cold molecular sources has recently been a
particularly active field of research
\cite{rotating-source,stark1,hfs-stark,chandler,elecfilter,fulton:243004}.
As with atoms, one of the aims is to produce quantum degenerate
gases, including those comprising strongly interacting electric
dipoles \cite{dipolarBEC,dipolarFermi,bdg02}.

For the purposes of loading traps, an ideal source would produce a
large flux of any atom or molecule at temperatures less than the
depth of the trap, T$_t$.  For most currently used trap
technologies, T$_t$ $\lesssim 1$~K.  For species amenable to laser
cooling, this temperature is within the capture range of a MOT
\cite{OriginalMOT}. For paramagnetic atoms and molecules, strong
magnetic traps can have depths exceeding 1K \cite{HMN04trap}.  For
polar molecules, electric field-based traps can reach similar
depths \cite{microwavetrap,bbc00}.

We describe here a simple, robust source that can operate with
nearly any atomic or molecular species, and which produces a beam
at high flux with translational and rotational temperatures near 1
K.  We demonstrate this source for both an atom (Na) and a polar
molecule (PbO), and discuss the different regimes of beam
formation.  Our analysis makes it possible to estimate and
optimize various characteristics of the beam source for general
use.  We
believe this provides an attractive alternative to many beam
techniques now in use.

%
%

A simple outline of the operation of our source is as follows.
Atoms or molecules of the desired species, \textbf{A}, are first
vaporized by laser ablation.  This produces $N$ particles of
\textbf{A} per pulse, at a high temperature $T_i$.
The ablation takes place inside a cryogenic cell filled with He
buffer gas at low temperature $T_b$ and density $n_{He}$.  After a
characteristic number of collisions $\cn$, the translational
temperature $T$ of \textbf{A} comes arbitrarily close to
equilibrium with the buffer gas, such that $T \approx
(1+\epsilon)T_b$ when $\cn = -\kappa ln(\epsilon T_b/T_i)$ \cite{primer}.
[Here $\kappa \equiv (m_A + m_{He})^2/(2m_A m_{He})$, and $m_{He}$
$(m_A)$ is the mass of He (\textbf{A})].
Rotational degrees of freedom are also cooled during these collisions.
By allowing both He and
\textbf{A} particles to exit the cell via a small hole, a beam of
\textbf{A} is formed.  The beam persists for a duration given by
the diffusion lifetime of \textbf{A} in the cell, which is limited
by sticking of \textbf{A} particles upon contact with the cell
walls.

The number of cold particles of \textbf{A} in the beam is
determined by both $n_{He}$ and the cell geometry.  During
thermalization, a particle of \textbf{A} typically travels a
distance $R \approx \cn /(n_{He} \sigma_t)$, where $\sigma_t$ is a
thermally-averaged cross section for elastic
collisions. Hence, for a cell with distance $R_{h}$ from the
ablation point to the hole, the particles of \textbf{A} will be
efficiently thermalized before exiting the hole only if $R < R_h$.
In addition, the purely geometric probability for a particle of
\textbf{A} to escape in the beam is governed by the ratio $d/R_h$,
where $d$ is the diameter of the exit hole.

The forward velocity, $v_f$, of the thermalized beam of \textbf{A}
particles also depends on both $n_{He}$ and $d$.  Specifically,
$v_f$ is determined by the ratio $d/\lambda$, where $\lambda =
1/(n_{He} \sigma_c)$ is the mean free path of \textbf{A} particles
in the cell; here $\sigma_c$ is the elastic cross-section for cold
\textbf{A}-He collisions.  In the effusive limit $(\lambda \gg
d)$, $v_f$ will be given approximately by the thermal velocity of
cold \textbf{A} particles, i.e. $v_f \approx v_{A} \equiv
\sqrt{2k_BT_b/m_A}$.  By contrast, when $\lambda \ll d$ the
\textbf{A} particles will become entrained in the outward flow of
He, so that $v_f \approx v_{He} \equiv \sqrt{2 k_B T_b/m_{He}}$.
Since $m_A \gg m_{He}$ for most species of interest, $v_f$ is much
smaller in the effusive limit than for an entrained beam.

Note that the conditions for efficient thermalization and for a
slow beam are in conflict.  Thermalization is most efficient for
$n_{He}$ above a threshold value, but effusive flow demands that
$n_{He}$ be less than a typically different threshold.  The
highest flux of cold, slow \textbf{A} particles is obtained when
$R_h$ and $d$ are chosen so that these thresholds in buffer-gas
density coincide (namely, when $R_h/d \approx
\cn\sigma_c/\sigma_t$), and $n_{He}$ is set at this common
threshold value ($n_{He}^{-1} = d\sigma_c = R_h\sigma_t$).

%
%
\begin{figure}
  \includegraphics[scale=1]{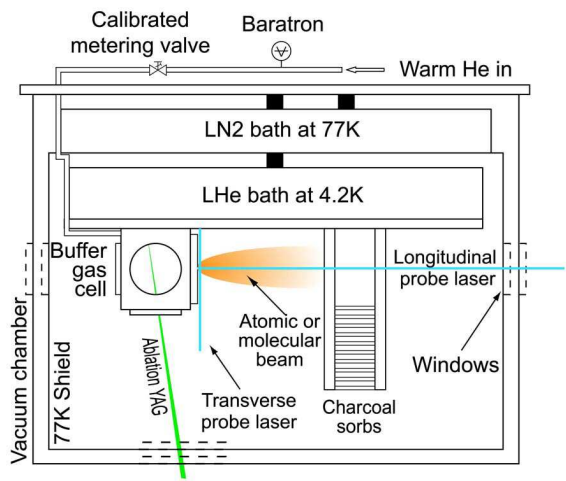}\\
  \caption{A schematic of the beam apparatus.  When detecting fluorescence, a lens (not shown) collimates a fraction of the fluorescence light and directs it out a window to a PMT.}\label{completebeam}
\end{figure}

A schematic of the beam setup is given in Fig. \ref{completebeam}.
The buffer gas cell is a brass box $\sim \!\! 10$~cm on edge.  The
cell is mounted in vacuum, with the top face attached to the cold
plate of a liquid He cryostat.  An $0.8$~mm long exit hole with
$d=3$~mm is centered on one side face. The bottom and other side
faces are covered with windows for optical access.  Several
ablation targets are mounted on the top face at $R_h\approx6$~cm.
The ablation light consists of laser pulses of $\sim \! 5$ ns
duration, with energy $\sim \! 15$ mJ, focused to a spot size
$\lesssim1$~mm, at a wavelength of 532 nm. The ablation laser is
typically fired at 10~Hz repetition rate. We produce Na atoms with
sodium metal or NaCl targets, and PbO molecules with a vacuum
hot-pressed PbO target. With our ablation conditions, typically
$T_i \approx 1000$ K for both species.

Buffer gas continuously flows into the cell through a narrow tube
which is thermally anchored to the cold plate. This ensures that
the buffer gas is at the temperature of the cold plate upon entry
to the cell.  The cell walls, and thus the He gas, are typically
at $T_b = 5$~K. Helium in the cell must be replenished as it flows
continuously through the exit hole. A calibrated metering valve at
room temperature is used to control the flow into the cell, and
hence the density, of the buffer gas. We
determine $n_{He}$ to within a factor of 2.  Good vacuum is
maintained
in the beam region
 by means of a coconut charcoal sorption pump with a pumping speed
of $\sim \!\! 1000$~l~s$^{-1}$.

For a typical elastic cross-section $\sigma_c \approx 3 \times
10^{15}$~cm$^{-3}$, the crossover between effusive and entrained
flow of \textbf{A}\textemdash i.e., the condition $d =
\lambda$\textemdash occurs for $n_{He} = n_c \approx 10^{15}$
cm$^{-3}$. This should be compared to the density $n_{He} = n_t$
required for full thermalization of \textbf{A}
particles\textemdash i.e., such that $R \approx \cn /(n_{He}
\sigma_t) = R_h$. Assuming $\sigma_t \approx \sigma_c$, we find
that for Na in our cell, $n_t \approx n_c$. Thus it should be
anticipated that our cell is near the optimal geometry for
producing a maximal flux of slow, cold Na. By contrast, the larger
mass of PbO makes $\cn$ much larger than for Na, implying that our
cell geometry is not optimal for PbO.  We characterize the the
beam source for both species within a range of densities around
the anticipated optimal condition for Na, namely $n_{He} \approx
0.2-5 \times 10^{15}$~cm$^{-3}$.

The beam source is monitored using laser spectroscopy.  Doppler
shifts and widths of the spectra are used to determine beam
velocity profiles. Signal size and timing yield the column density
and particle dynamics.  To measure the longitudinal (transverse)
velocity profile, a probe laser beam is sent collinear with
(perpendicular to) the molecular beam.  For Na, a third probe beam
monitors atoms inside the cell.

We monitor Na atoms via absorption of a
probe laser tuned to the $3s_{1/2} \rightarrow 3p_{3/2}$
transition at wavelength $\lambda_{Na} = 589$ nm. The probe laser
frequency is continuously scanned over the entire absorption
profile (a range of $\sim 1$ GHz), at a rate of 1 kHz, resulting
in several complete spectral profiles of the Na atomic beam for
each ablation pulse.  The relatively slow scan of the laser
ensures that we are sensitive only to nearly-thermalized Na atoms;
hot atoms traverse the detection region before a scan is complete.

PbO is monitored via laser-induced fluorescence.
This probe laser is tuned to the X$(v''=0) \rightarrow $B$(v'=5)$
transition at $\lambda_{PbO,e} = 406$ nm. Fluorescence is detected
using a photomultiplier tube (PMT), with interference and colored
glass filters to selectively observe the B$(v'=5) \rightarrow$
X$(v''=4)$ transition at $\lambda_{PbO,f} = 460$ nm. A shutter
with an opening time of $\sim7$~ms is used to shield the PMT from
the initial glow following each ablation pulse. Again, this
technique ensures that only slow-moving molecules are detected.
The signal is averaged over several
(typically 10) shots of the ablation laser with the probe laser
frequency fixed.
A single spectral scan consists of frequency points separated by
10-30~MHz across a span of up to several GHz. Our quantitative
determination of molecule number from fluorescence counts includes
estimates of the detection efficiency and branching fraction for
the detected transition \cite{FCF}. We estimate the error in our absolute
number measurement of PbO to be less than a factor of 2.

\begin{figure}
   \includegraphics[angle=0,scale=1]{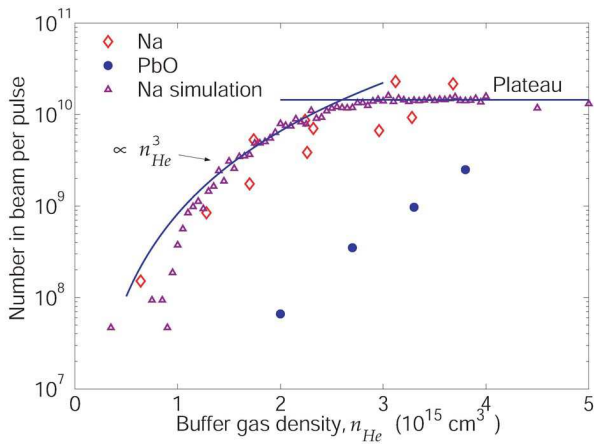}
   \caption{The number of cold \textbf{A} particles, $N_h$, that emerge in the
   beam as a function of $n_{He}$.  Curves with specific functional forms
   have been inserted to show the different scaling regimes.
   The intersection of the curves corresponds to the condition $R = R_h$, where a typical \textbf{A} particle
   is thermalized just as it reaches the hole.}
   \label{number}
\end{figure}

For Na, the in-cell probe beam was used to determine both $N$ and
$\sigma_c$.  We find $N_{Na} \approx 10^{14}$/pulse for both the
Na and NaCl targets. We measure diffusion lifetimes of $\tau \left[\mathrm{ms}\right]\approx 4\times 10^{-15}\times n_{He}\left[\mathrm{cm}^{-3}\right]$. From this we infer
$\sigma_c\approx 3\times10^{-15}$~cm$^2$.  For PbO, previous work has measured an ablation yield of $\approx 10^{12}$/pulse \cite{DimaPbO}, and our measurements with an in-cell probe indicate a comparable yield.

In Fig. \ref{number} we plot the number, $N_h$, of thermalized particles of species \textbf{A}
exiting the hole as a function of $n_{He}$.  We also show the results of a Monte Carlo
simulation of the beam formation process. The condition for full thermalization is
apparent in both the experimental and simulated data
for Na, while as expected we do not appear to reach the condition
of full thermalization for PbO.

In the Na data and simulations, we find that $N_h$ increases
rapidly (approximately $\propto n_{He}^3$) up to a critical value
of $n_{He}$, above which $N_h$ is roughly constant at its maximum
value $N_{h,max}$. The low-density scaling is consistent with a
simple picture in which \textbf{A} particles are distributed
uniformly over a volume of characteristic length $L_t \approx \cn
\lambda$, when they have thermalized to near $T_b$.  (This broad
distribution arises, in our simulations of the thermalization
process, from the spreads both in the number of collisions
required to thermalize and in the free path between collisions.)
At high density, the condition $N_h \simeq N_{h,max}$ arises in
our simulations when $R < R_h$ for essentially all \textbf{A}
particles.
In this regime, the fraction of \textbf{A} particles escaping,
$f_{max} = N_{h,max}/N$, is given roughly by the area of the hole
to the area of a hemisphere at radius $R_h$, i.e., $f_{max}
\approx d_h^2/(8 R_{h}^2)$.  In our geometry, $f_{max} \approx 3
\times 10^{-4}$.

The simulated beam data in Fig. \ref{number} is scaled to match
the experimental Na data by adjusting the values of $N_{Na}$ and
$\sigma_t$.  The resulting value, $N_{Na} \approx 5\times10^{13}$,
is in reasonable agreement with the determination from the in-cell
probe. The fitted thermalization cross-section, $\sigma_t \sim
1\times 10^{-15}$~cm$^2$, is somewhat smaller than $\sigma_c$.
This is reasonable, since elastic cross sections are typically
smaller at higher collision energies\cite{landau}.

\begin{figure}
   \includegraphics[angle=0,scale=1]{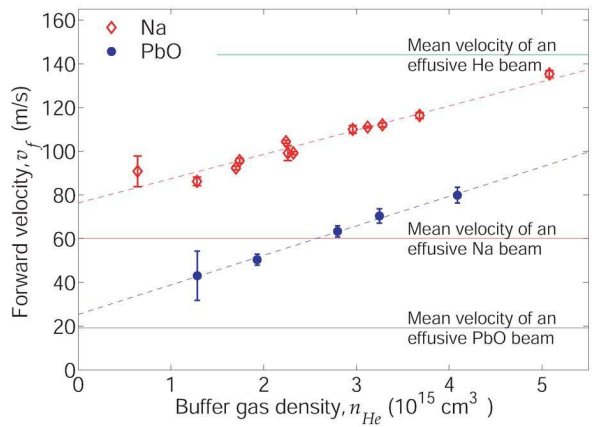}
   \caption{PbO and Na beam mean forward velocities, $v_f$,
   as a function of $n_{He}$.  Extrapolation of the data to zero buffer gas density is illustrated by best-fit lines (dashed).}
   \label{velocities}
\end{figure}

Fig. \ref{velocities} shows the average forward velocity, $v_f$,
of the beams of \textbf{A} particles as $n_{He}$ is varied. For
both Na and PbO, the data show a nearly linear increase of $v_f$
with $n_{He}$, with the velocity of the lighter species always
larger. This behavior is consistent with the following simple
picture. A slowly-moving particle of \textbf{A} takes a time $T_e$
to exit the hole, where $T_e \sim d/v_A$. During this time, it
undergoes $N_e$ collisions with fast, primarily forward-moving He
atoms, where $N_e \sim n_{He} \sigma_c v_{He} T_e$. Each collision
imparts a momentum transfer $\Delta p_A \sim m_{He} v_{He}$.  This
results in a net velocity boost $\Delta v_A$, given by $\Delta v_A
\sim v_A d/\lambda \propto n_{He}$. This picture should be roughly
valid for densities below the regime of full entrainment, where
$v_f \sim v_{He}$.  The velocities we measure for Na are
approximately reproduced by modeling of the beam formation
process with our measured value of $\sigma_c$.

This picture also predicts that the behavior of $v_f$, when
extrapolated to $n_{He} = 0$, should yield the velocity of an
effusive beam of \textbf{A} particles at temperature $T_b$.  To
make this comparison, it is critical to note that our detection
technique is sensitive to molecules within a roughly cylindrical
volume, of diameter $D_d$ and with length $L$ extending from the
exit hole.  Under our conditions, where $D_d \sim d < L$, it can be
shown that an effusive beam will exhibit a velocity
distribution close to $f(v) \propto v^2 e^{-m_Av^2/(2k_BT_b)}$ and a mean
velocity $\bar{v}_{eff} \approx 1.13 v_A$.  Our extrapolated data
is within $\sim \! 25 \%$ of this prediction for both species.

\begin{figure}
   \includegraphics[scale=1,angle=0]{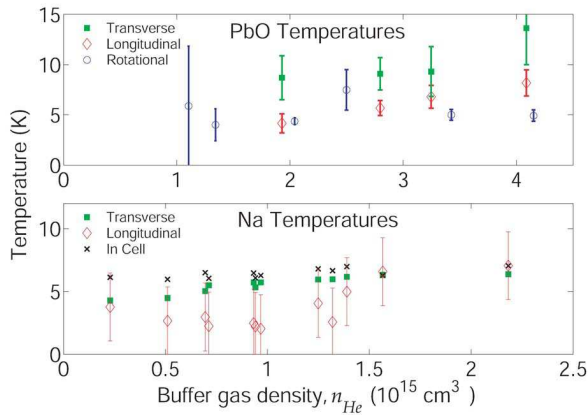}
   \caption{PbO and Na temperatures as a function of $n_{He}$ as determined by fits to spectral profiles. ``Transverse'', ``longitudinal'', and ``in cell'' refer to the location of the probe laser (see Fig. \ref{completebeam}). Typical errors for the transverse and in cell Na temperatures are comparable to the marker size.}
   \label{temps}
\end{figure}

Fig. \ref{temps} shows the temperature of the beams vs. $n_{He}$,
as measured by the velocity spreads in the longitudinal and
transverse directions, as well as the rotational population
distribution (for PbO).  For transverse temperature we fit to a
distribution of the form $f_t(v)\propto e^{-m_A v^2/(2 k_B T)}$.
For longitudinal temperature we use $f_l(v)\propto e^{-m_A
(v-v_l)^2/(2 k_B T)}$.  Fits of the longitudinal data to the
effusive distribution $f_{l,eff}(v) \propto v^2 e^{-m_s v^2/(2 k_B
T)}$ were poor, consistent with the partial entrainment of
\textbf{A} in the helium flow. For the rotational temperature, we
use Clebsch-Gordan and H\"{o}nl-London factors to determine the
relationship between fluorescence intensity and initial state
population for various rotational lines.  The ratio of initial
state populations for 2-3 rotational lines determine the
temperature corresponding to a Boltzmann distribution.  Our data
indicates complete thermalization of all detected particles.  Note
that we observe no additional cooling below $T_b$, as would be
expected for entrainment in a fully supersonic He flow.
%
%
%

The beam source described here can be readily adapted to the needs
of a wide range of experiments.  For example, it can be used for a
wide variety of species with performance similar to that described
here.  The total flux depends linearly on the ablation yield $N$,
which for any given target is difficult to predict \textit{a
priori}.  However, for nearly every species we have tried (here
and in many related experiments), it has been possible to achieve
large values of $N$ by a suitable choice of precursor material.
For example, under operating conditions similar to those used
here, we have obtained $N = 10^{12}\! -\! 10^{14}$ for a variety
of metal atoms and $N = 10^{11}\! -\! 10^{13}$ for many species of
diatomic molecules (including radicals)
\cite{Eupaper,bretislavcah,bretislavCaF,rareEarth}.  In
addition--subject to limitations of cooling power or gas load--the
source could be run at lower temperatures, higher repetition
rates, or with higher extraction efficiency (e.g., by using
several separated exit holes).  Our discussion of the beam
formation mechanism makes it straightforward to determine the
effects of such changes.

A key possible improvement to the source would be the addition of
a guide, either magnetic (for paramagnetic species) or electric
(for polar molecules).  In both cases, He is unaffected by the
guide potential and will exit through the sides of the guide,
allowing extraction of the \textbf{A} beam into a region of
ultra-high vacuum.  Under the conditions described here, this beam
source could be used to load a peak flux of up to
$2\times10^{11}$~s$^{-1}$ Na atoms into a simple permanent magnet
guide such as the type described in Ref. \cite{abrahamMagfilter}.
An electrostatic guide such as that described in Ref.
\cite{elecfilter} could be loaded with a peak flux of $\sim
10^9$~s$^{-1}$ PbO molecules in the J=1 rotational state, adequate
for loading into a microwave \cite{microwavetrap} or electrostatic
\cite{ammoniatrap} trap.
The high fluxes from our source could
result in substantial improvements in atomic and molecular
trapping experiments that depend on large initial numbers.

We acknowledge support from the National Science Foundation under Grant Nos. DMR-0325580 and PHY-0071311, the Army Research Office, the W.M. Keck Foundation, and the David and Lucile Packard Foundation.
\bibliographystyle{apsrev}

\end{document}